\documentclass[showpacs,numbers,amsmath,amssymb,preprint,prl,twocolumn]{revtex4-1}
\usepackage{graphicx}
\usepackage{dcolumn}
\usepackage{bm}

\usepackage{graphics}
 \usepackage{graphicx}
 \usepackage{epsfig}

\usepackage{amssymb}
\usepackage{amsmath}
\usepackage[normalem]{ulem}

\usepackage{color}
\definecolor{darkgreen}{rgb}{0,.6,0}

\begin{document}


\preprint{}
\title{Microstructure evolution and heterogeneous nucleation in ternary Al-Cu-Ni alloys}
\author{Julia Kundin}
\email{julia.kundin@uni-bayreuth.de}
\author{E. Pogorelov}
\author{H. Emmerich}
\affiliation{Material and Process Simulation (MPS), University Bayreuth, 95448 Bayreuth, Germany}

\date{\today}

\begin{abstract}
The simulations of the solidification of ternary Al-Cu-Ni alloys by means of a general multi-phase-field model for an arbitrary number of phases reveal that the real microstructure can be generated by coupling the real thermodynamic parameters of the phases and the evolution equations. 
The stability requirements on individual interfaces for model functions guarantee an absence of "ghost" phases  in a $n$-dimensional phase-field space. The special constructed thermal noise terms disturb the stability and can produce the heterogeneous nucleation of product phases in accordance to the energetic and concentration conditions. Of particular interest is that in triple points the nucleation of the forth phase occurs without additional noise. Another observation is the growth of the eutectic-like or peritectic-like structure in various alloys. 
\end{abstract}
\maketitle
The modeling of the microstructure can not only predict how the amounts of the phases and their composition vary with temperature or chemistry for a given alloy but also define the size and distribution of structure components.
The evolution of the realistic microstructure of multi-component alloys was studied in a wide range of the works by means of the multi-phase-field approaches developed in the recent years. These approaches can be divided into two main group: models based on multi-phase concept of Steinbach \cite{Steinbach96} and models, which exploit the Lagrange multiplier formalism \cite{Garcke04,Folch05}. The main works in this area are the phase-field modeling of eutectic \cite{Nestler00,Apel02,Kim04,Nestler05,Folch05} and peritectic solidification \cite{Tiaden98,Lo03,Choudhury10}. However, the investigations were limited by the three-phase transformations and the four-phase transformation reactions are not fully covered.
 
The focus of the paper is the investigation of new effects in ternary alloys, where the four-phase reactions can occur.  For our study we chose the Al-reach corner of the phase diagram of the Al-Cu-Ni system as a ternary model system, first because it has the advanced thermodynamic and phase equilibria data and second because the previous knowledge in the nucleation and the microstructure formation on the binary edge systems Al-Ni \cite{Kundin11,Kundin12,Kundin12b} and Al-Cu \cite{Boettger09} can be directly included. Moreover, many experimental studies of the microstructure in ternary alloys have been reported in the literature \cite{Wang01,Gamer03,Yan12} and there is the need for the theoretical comprehension of the various structure phenomena.

We simulate the microstructure evolution and the following nucleation effects by means of the general phase-field model for multi-component systems in $n$-phase-field space \cite{Pogorelov13}. In this model the method of Lagrange multipliers and the idea of flatness and stability requirements suggested in the model of Folch and Plapp \cite{Folch05} were implemented to construct the special phase-field model functions. An interface noise is added to the evolution equations for the phase fields represented by Langevin
forces in accordance to the results of fluctuation theory described in Refs. \cite{Karma99,Changsheng05,Bronchart08} to simulate the heterogeneous nucleation on the macroscale.

In the model our physical system is fully described by $n$ phase fields $ p_{i}\in[0,1], i=1,2...n$ and $N$ chemical concentration fields $c^A \in[0,1], A=1,2...N$. 

The total free energy  functional  of the system is given as
\begin{eqnarray}
F(\bm{p},\bm{c})&=&\int [\frac{K}{2}\sum_i (\nabla p_i)^2+H \sum_{i}f_{b,i}(p_i)\nonumber\\
& +&f_{c}(\bm{p},\bm{c},T)]d\bm{r}.
\end{eqnarray}
Here constants are defined as $K=W\sigma/a_1$ and $H=\sigma/(Wa_1)$, where $W$ is the interface width, $\sigma$ is the interface energy over all dual interfaces, $a_1$ is a numeric constant, $\bm{c}$ is the mixture composition vector, T is the temperature. Then $f_{b,i}$ are barrier functions of phases and $f_c$  is the chemical part of the free energy density.

For non-conserved phase fields we can write an equation of motion using the Ginsburg-Landau equation with Lagrange multiplier \cite{Folch05}
\begin{equation}
\tau(\bm{p})\frac{\partial p_{i}}{\partial t}=-\frac{1}{H}\frac{\delta F}{\delta p_{i}}\biggm|_{\sum p_{j}=1}=-\frac{1}{H}\biggl(\frac{\delta F}{\delta p_{i}}-\frac{1}{n}\sum_{j}^n\frac{\delta F}{\delta p_{j}}\biggr) \label{EqLagrangeMulti}
\end{equation}
that ensures $\sum_i \partial p_i/\partial t = 0$.
Here $\tau(\bm{p})$ is a system relaxation time, depending on the local values of the phase fields and the value of the driving forces \cite{Karma98,Karma01}.

The corresponding barrier functions are chosen as $f_{b,i}=p_i^2(1-p_i)^2$. The model functions $g_i(\bm{p})$ in the chemical free energy $f_c$ are constructed to fulfill the stability and flatness requirements in general $n$-dimensional space and have the form \cite{Pogorelov13}
\begin{equation}
g_i(\bm{p})=\frac{p_i^2}{2}  \biggl( 15(1-p_i)(1-p_i-\sum_{j\ne i} p_j^2)+p_i(5-3p_i^2)\biggr).
\end{equation}

The mixture chemical free energy of a multi-component system can be written using the second order Taylor expansion around the minimal composition
\begin{eqnarray}
f_c&=&f_c^{m} + \sum_{A,B}^n\frac{X^{AB}}{2}\left(c^A-c^{A,m}\right)\left(c^B-c^{B,m}\right),
\label{LagrangeMulti}
\end{eqnarray}
where $f_c^{m}=\sum_{i}^n  B_{i}g_{i}$ is the mixture chemical free energy in the minimum with $ B_{i}(T)$ being the minimal chemical free energies of phases, $c^{A}$ are components of the mixture composition vector, $c^{A,m}=\sum_{i}^nA_{i}^Ag_{i}$ are components of the mixture minimal composition vector with  $A_{i}^A$ being the minimal concentrations, which are defined in the minimum of the chemical free energies of phases.
Then $X^{AB}$ are components of a mixture thermodynamic factor matrix, which is defined by the thermodynamic factor matrix of phases $\mathbf{\hat{X}_{i}}$ as \cite{Kundin12b} $\mathbf{\hat{X}}^{-1}=\sum_{i}^n \mathbf{\hat{X}_{i}}^{-1}g_{i}$.

We will also use mixture diffusion potentials as variables in the model equations
\begin{eqnarray}
\mu^A= \sum_{B}^NX^{AB}\left(c^B-c^{B,m}\right).  \label{TaylorMuMulti}
\end{eqnarray}
The chemical free energy~(\ref{LagrangeMulti}) gives the thermodynamic driving forces and a set of phase evolution equations has the form
\begin{eqnarray}
\tau(\bm{p})\frac{\partial p_i}{\partial t}&=&
W^2 \biggl(\nabla^2 p_i-\frac{1}{n}\sum_k^n \nabla^2 p_k\biggr)\nonumber\\
&-&\biggl(\frac{\partial f_b}{\partial p_i}-\frac{1}{n}\sum_k^n \frac{\partial f_b}{\partial p_k}\biggr)\nonumber\\
&+&\frac{1}{H}\sum_j^n\frac{\partial g_{j}}{\partial p_i}\biggm|_{\sum p_k=1} \Omega_j+\sum_{j \ne i}^n\xi_{ij},\label{PhasefieldEq0}
\end{eqnarray}
where we use the grand potentials of phases $\Omega_j=\sum_A^N\mu^{A}A_{j}^A-B_{j}$ for the sake of compactness.

The terms $\xi_{ij}$ represent the thermal nucleation noise and are constructed as
\begin{eqnarray}
\xi_{ij}&=&r\xi_0\frac{RT}{H^2}\frac{15}{(n-1)}\left(p_j^2(1-p_j)( p_L- p_i)\Omega_j \right.\nonumber\\
&-& \left. p_L^2(1-p_L)( p_j- p_i)\Omega_L\right), \label{noise}
\end{eqnarray}
where  $r\in [-0.5,0.5]$ is a random number,  $\xi_0$ gives the magnitude of the fluctuations, $p_L$ is the liquid phase field and $p_i$, $p_j$ are solid phase fields. The noise terms produce the fluctuations of a phase $i$ on a $j$/liquid interface with an amplitude proportional to the driving force $\Omega_j-\Omega_L$ and back proportional to the surface energy. It can be shown that the noise terms (\ref{noise}) depend on the undercooling and the surface tension in consistence with the nucleation theory. Indeed, on a dual $j$/liquid interface in our four-phase system the noise terms for the phase $i$ reduce to $\xi_{ij}=r\xi_0\,\frac{RT}{H^2}5p_L^2(1-p_L)^2(\Omega_j-\Omega_L)$. This is proportional to the probability of the nucleation for the phase $i$ $, \exp(-\frac{\Delta G}{RT})\approx\frac{RT}{\Delta G}$, with a nucleation barrier for the 2D system, $\Delta G\approx\frac{\sigma^2}{\Delta \Omega}$.

The diffusion equations for all components have the following form
\begin{eqnarray}
\frac{\partial c^A}{\partial t}&=&\nabla \cdot \left[\sum_B^n M^{AB}(\bm{p})\nabla \mu^B-\bm{J}_{at}^A(\bm{p})\right],\label{B4}
\end{eqnarray}
where $M^{AB}$ are the components of the mobility matrix $ \mathbf{\hat{M}}=\mathbf{\hat{D}}\cdot \mathbf{\hat{X}}^{-1}$. The component of the diffusion matrix are defined as $D^{AB}(\bm{p})=\sum_i^N D_i^{AB}g_i$ with $D_i^{AB}$ being the terms of the diffusion matrix in a phase $i$.
The values $\bm{J}_{at}^A$ are the anti-trapping currents for all components.

The scaled material and model parameters used in the simulations are presented in
Table~\ref{Tabel1}. The system time scale was chosen as $1 \times 10^{-3}$ s, the system length scale as $1.3\times 10^{-8}$ m and the energy scale had the order of the  thermodynamic factor in a liquid for all elements $E_0=2 \times 10^{6}$ J/mol-at. In the simulations we neglected the cross terms in the diffusion and thermodynamic factor matrices.

\begin{table}
\caption{Scaled material and phase-field model parameters used in the simulation.}
\label{Tabel1}
\begin{center}
\begin{tabular}{ l c}
\\ \hline  \hline
Parameter  & Value \\  \hline
$\tilde{\tau}_{ij}=\tilde{\tau}$ (relaxation time) & $1$\\
$\tilde{\Delta x}$ (grid discretization size)	& $1$\\
$\tilde{\Delta t}$ (time step) & $0.025$\\
$\tilde{W}$ (interface width) & $1.2 $\\
$\tilde{D}_L^{Ni}$ (diffusion in liquid phase) & $1$\\
$\tilde{D}_L^{Cu}$ (diffusion in liquid phase) & $0.8$\\
$\tilde{D}_S$ (diffusion in solid phase) & $0.01 D_L$\\
$\tilde{H}$ (scaled surface energy) & 0.14\\
$\xi_0$ (amplitude of noise) & $0\div0.6$\\
\hline \hline
\end{tabular}
\end{center}
\end{table}

For the numerical tests we chose alloy 1 with an initial concentration of 4 at$\%$Cu-11 at$\%$Ni and alloy 2 with  6 at$\%$Cu-19 at$\%$Ni. The main interest of the study is the four-phase reaction at  604$^\circ$C:
\begin{eqnarray}
\text{L} + \text{Ni}\text{Al}_3(\alpha)\rightarrow \text{Ni}_2\text{Al}_3(\beta) + \text{(Al)}(\gamma).\label{Reaction}
\end{eqnarray}

 Equations (\ref{PhasefieldEq0}) and (\ref{B4}) were solved numerically using the Euler method in the cubic 2D simulation box of size $1000\,\Delta x$ and $500\,\Delta x$ with periodic boundary conditions. To verify the microstructure formation we have also prepared 3D simulations in a box of size $100\,\Delta x$.

 The microstructure of alloy 1 was modeled in two tests at a constant temperature of 575$^\circ$C. The used thermodynamic parameters are listed in Table 2. The parameters $A_i^A$ for $\alpha$ and $\beta$ phases are chosen according to the equilibrium phase diagram. In test 1 the concentrations for the $\gamma$ phase are chosen two times larger than according the phase diagram. This is for the purpose to proof the influence of this parameter on the microstructure. We started with the growth of the initial crystals of $\alpha$ phase. After 1200 time steps the nuclei of the $\beta$ and $\gamma$ phase were inserted randomly on the $\alpha$/liquid interface. In Figure 1 the microstructure of alloy 1 for the test 1 is shown. The $\gamma$ phase starts to grow between the crystals of the  $\alpha$ and $\beta$ phase only if the  fraction of the $\beta$ phase is larger then 20$\%$. In Figure 2 the microstructure of alloy 1 for test 2 is shown. The coupling eutectic-like growth of the $\beta$ and $\gamma$ phase is 
observed. The crystals of the $\gamma$ phase precipitate along with $\beta$ during the four-phase peritectic reaction (\ref{Reaction}). The structure is similar to the experimental microstructure observed in Al-Cu-Sn and Al-Cu-Ni-Mg alloys \cite{Yan12,Belova05}. The basis for the occurrence of the lamellar structure is that the both product phases, $\beta$ and $\gamma$, have similar Gibbs free energies and symmetric concentration parameters. In Figure 2(a) the simulations are carried out without the thermal noise. The eutectic-like structure forms by the overgrowing of one phase over other one. In Figure 2(b) the amplitude of the noise was chosen with $\xi_0$=0.45. The first interesting observation is that the 
noise refines the eutectic structure such that the lamellar thickness decreases by 2-5 times.  The crystals of $\gamma$ phase precipitate on the $\beta$ phase and the new crystals of $\beta$ phase form on the $\gamma$/liquid boundaries as shown in Figure 3(a). The both phases serve as nucleants for each other. 

\begin{figure}
\label{Fig1}
\begin{centering}
\begin{tabular}{ll}
\includegraphics[scale=0.2]{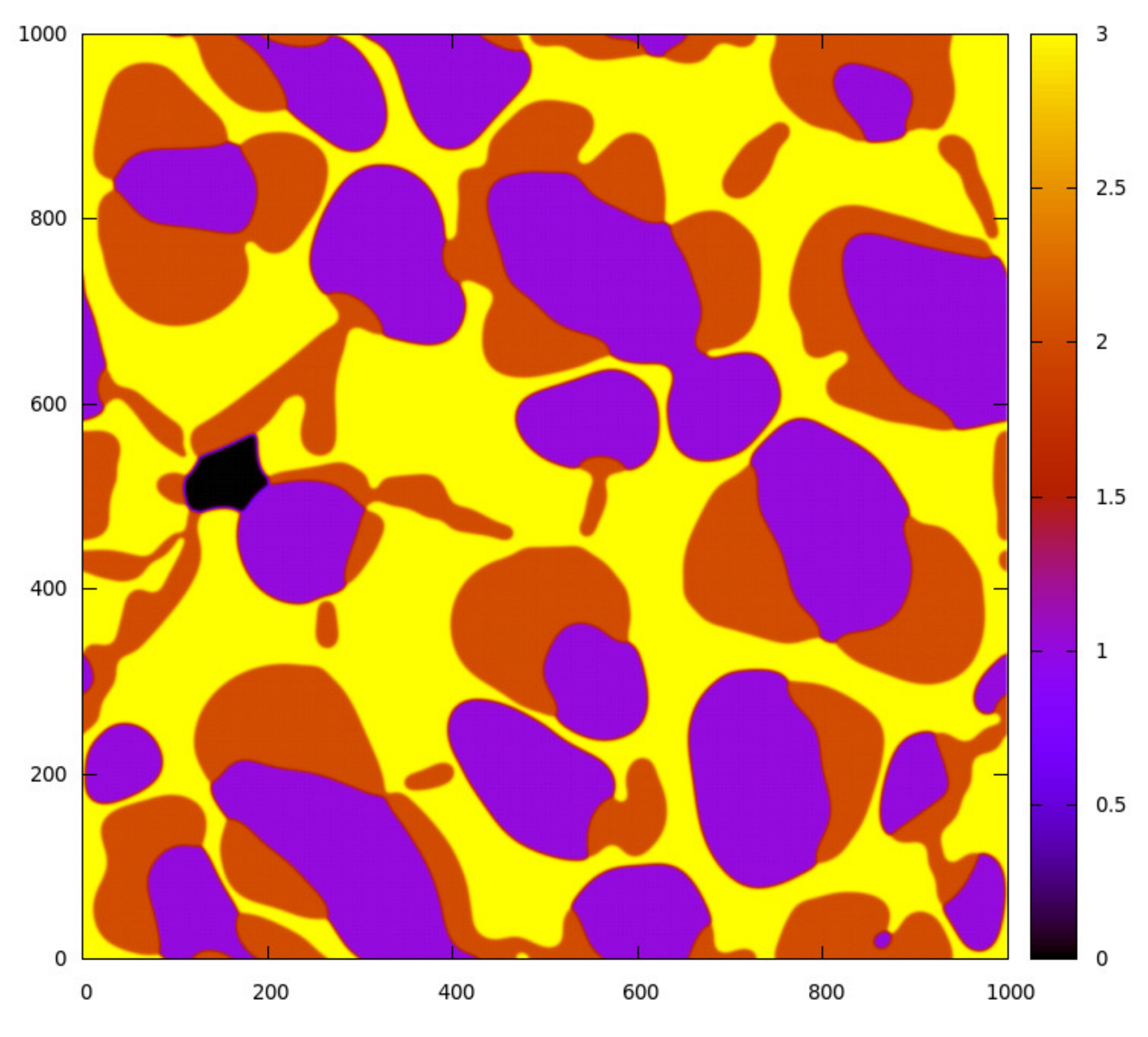}&
\includegraphics[scale=0.2]{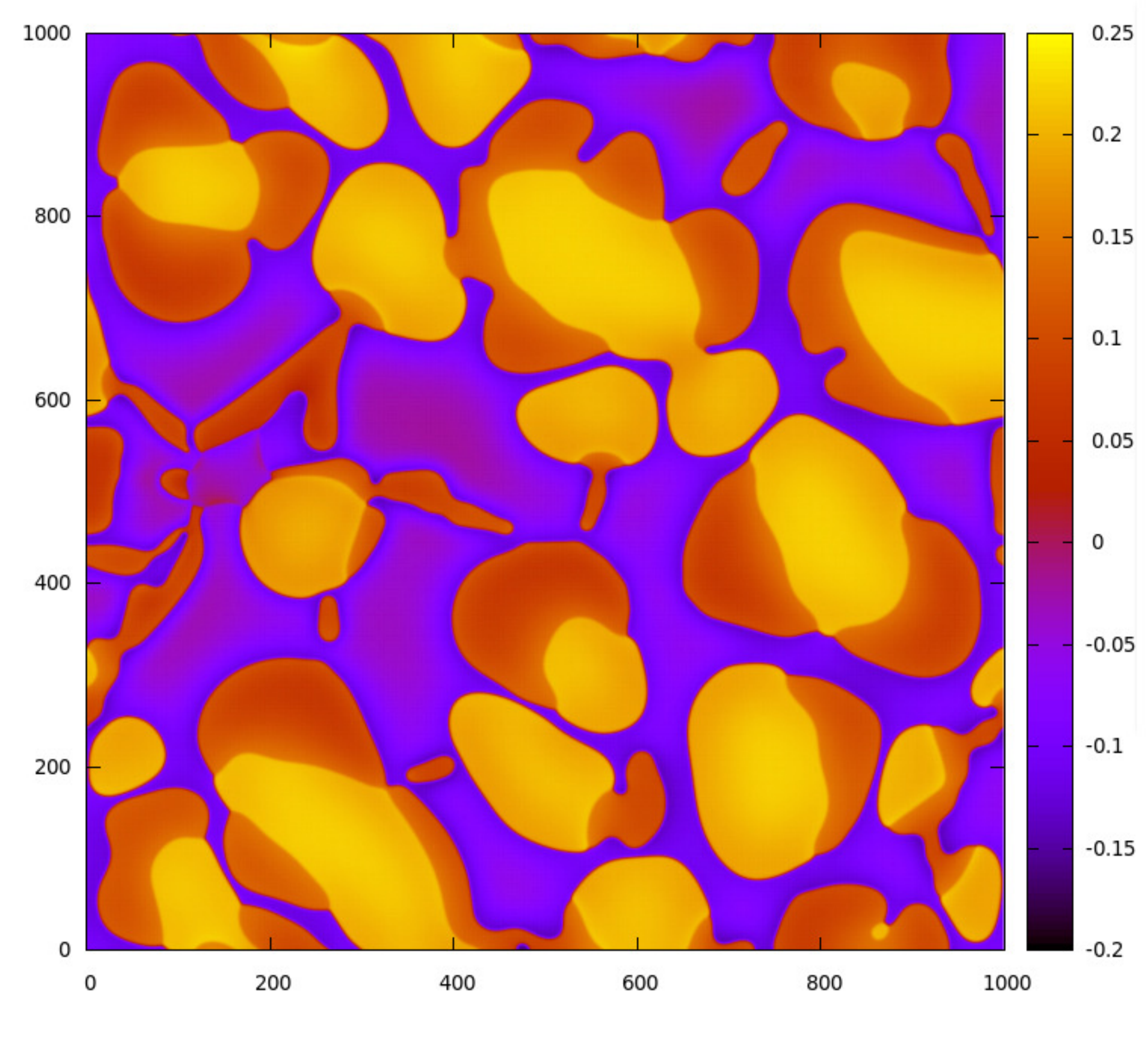}\\
(a)&(b)
\end{tabular}
\caption{Simulated microstructure in alloy 1 for test 1 (a) and the corresponding concentration field of Ni (b). Purple, red and yellow areas indicate $\alpha$, $\beta$ and $\gamma$ phases, respectively. Black area indicate the rest of the liquid phase. }
\end{centering}
\end{figure}

\begin{figure}
\label{Fig2}
\begin{centering}
\begin{tabular}{ll}
\includegraphics[scale=0.16]{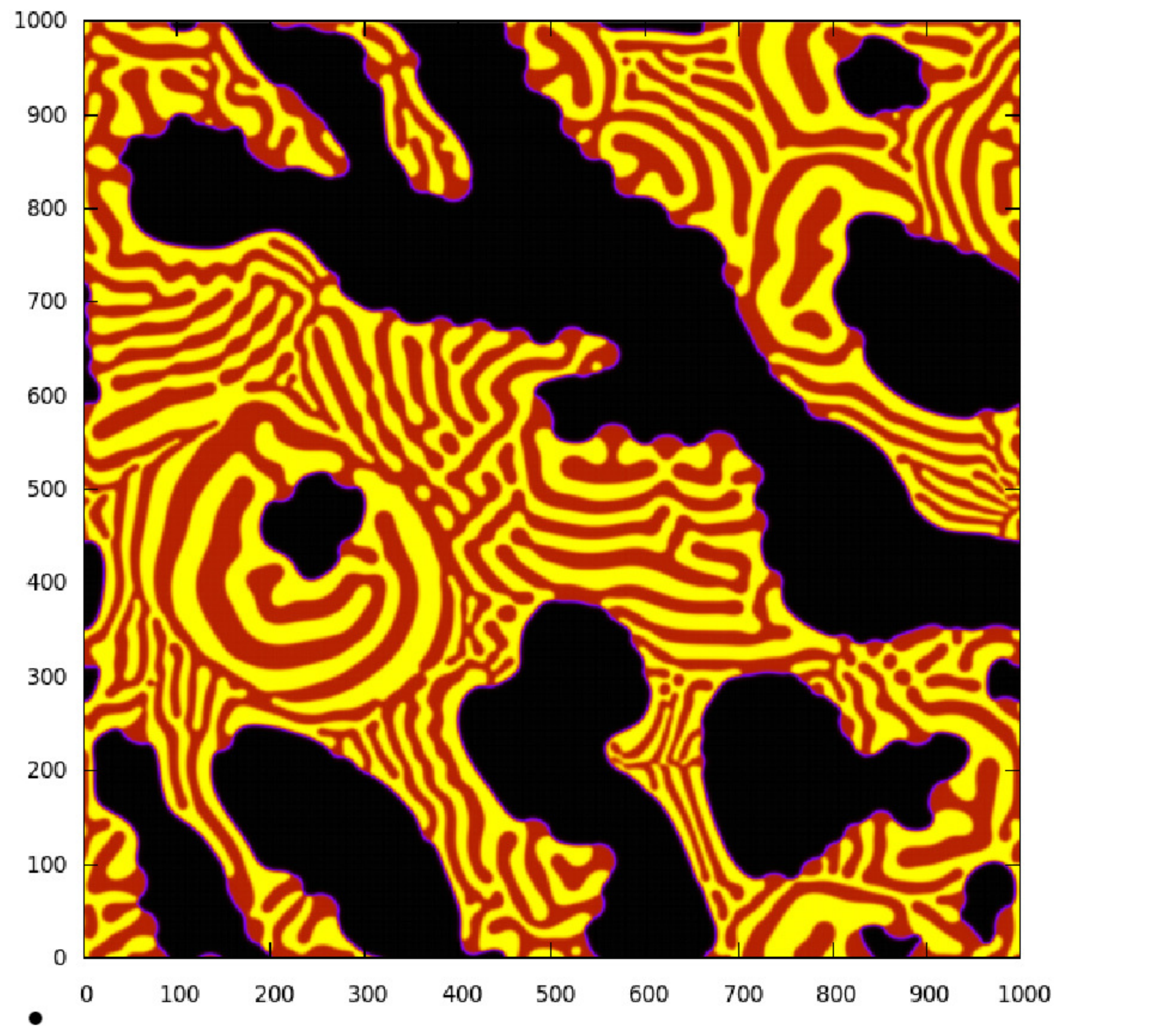}&
\includegraphics[scale=0.16]{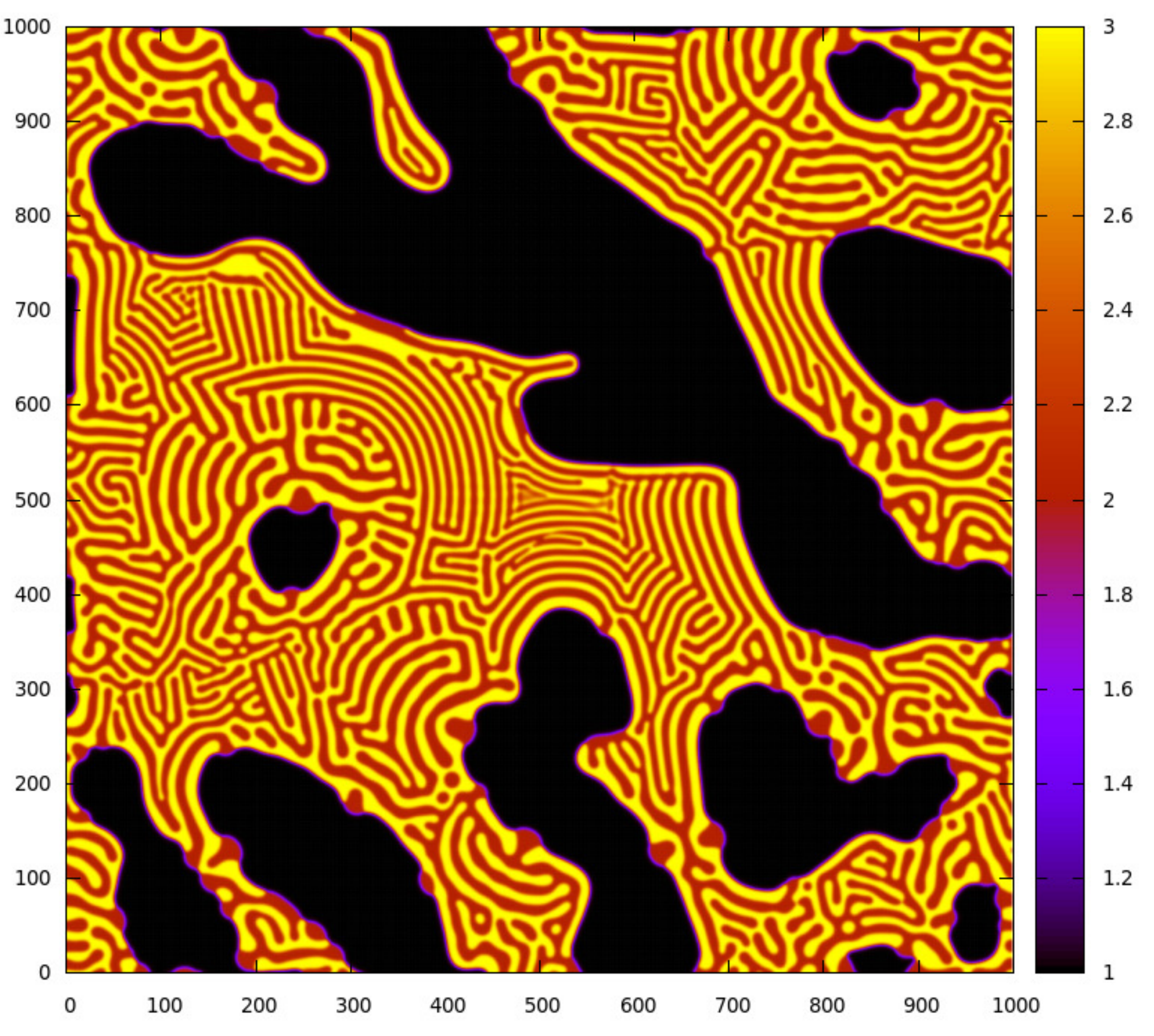}\\
(a)&(b)
\end{tabular}
\caption{Simulated microstructure of alloy 1 for test 2 without (a) and with (b) noise. Black, red and yellow areas indicate $\alpha$, $\beta$ and $\gamma$ phases, respectively.  }
\end{centering}
\end{figure}

\begin{table}
 \begin{center}
\caption{Thermodynamic parameters of phases in alloys 1 and 2 used in the simulations. $\tilde{A}_i =A_{i}-A_{L}$, $\tilde{B}_i =\frac{B_{i}-B_{L}}{E_0}$, $\tilde{X}_i =\frac{X_{i}}{E_0}$ }
  \begin{tabular}{ccccccccc}
\hline
 alloy&phase &  $\tilde{A}^{Ni}_{i}$ &  $\tilde{A}^{Cu}_{i}$ &   $\tilde{B}_{i}$  & $\tilde{X}_{i}^{Ni}$ & $\tilde{X}_{i}^{Cu}$ & $\tilde{X}_{i}^{NiCu}$\\
 \hline  
1&$\alpha $  & 0.24   &0.0    & -0.012 & 10  &10&0\\ 
 &$\beta $  &  0.17  & 0.20 &-0.025  & 5  &5&0\\
&$\gamma $ test 1 &  -0.17 & -0.17  &-0.025& 1 & 1&0  \\
&$\gamma $ test 2 &  -0.08 & -0.08  &-0.006& 1 & 1 &0 \\\hline
2&$\alpha $  & 0.22   &-0.007     & Fig.~2 & 10 & 10&0 \\ 
 &$\beta $  &  0.18 & 0.19  & Fig.~2  & 5& 5&0   \\
&$\gamma $  & -0.08 &-0.08& Fig.~2  & 1 & 1 &0 \\\hline
  \end{tabular}
 \end{center}
\label{Table2}
\end{table}

In the simulation we can also observe that new crystals of the $\gamma$ phase form in the triple points of the phases. This occurs with and without noise terms. Furthermore, Figure 3(b) shows the 3D view of the nucleation of the $\gamma$ phase on a triple line with the same mixture of $\alpha$, $\beta$ and liquid phases.

 For alloy 2 the simulations were carried out with the equilibrium parameters presented in Figure 4 and Table 2. During the simulation the temperature decreases linear with the cooling velocity, $v_c=5$ K/s. Note that the cooling rate strongly influences the phase fraction evolution.
 After  575$^\circ$C the temperature is assumed to be constant due to the latent heat extraction during the four-phase reaction. We started with the growth of initial crystals of the $\beta$ phase. At  604$^\circ$C nuclei of the $\alpha$ phase were inserted at random sites on the solid/liquid interface of the $\beta$ phase and after any steps the crystals of the $\gamma$ phase nucleate spontaneously and grow along with the $\beta$ phase. The nucleation occurs in triple points of $\alpha$, $\beta$ and liquid phases  as shown in Figure 3(b,c).

\begin{figure}
\label{Fig3}
\begin{centering}
\begin{tabular}{lll}
\includegraphics[scale=0.25]{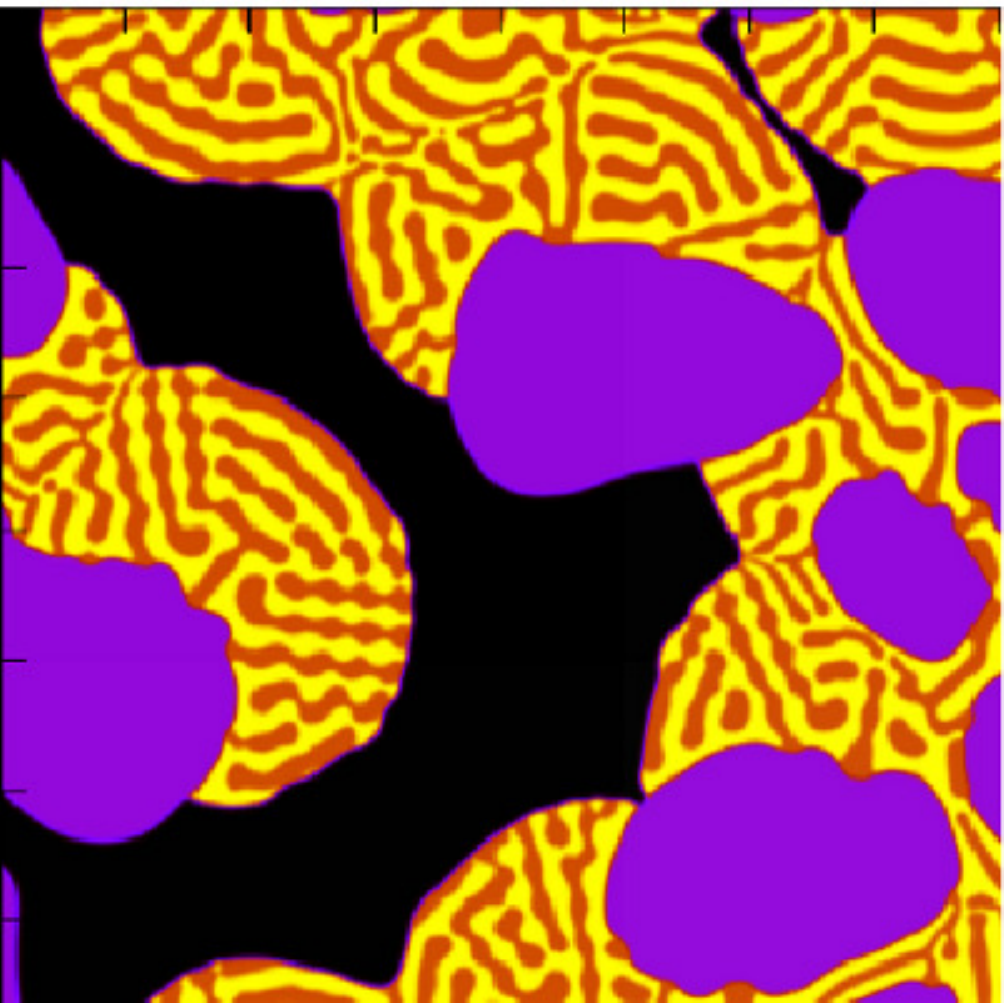}&
\includegraphics[scale=0.25]{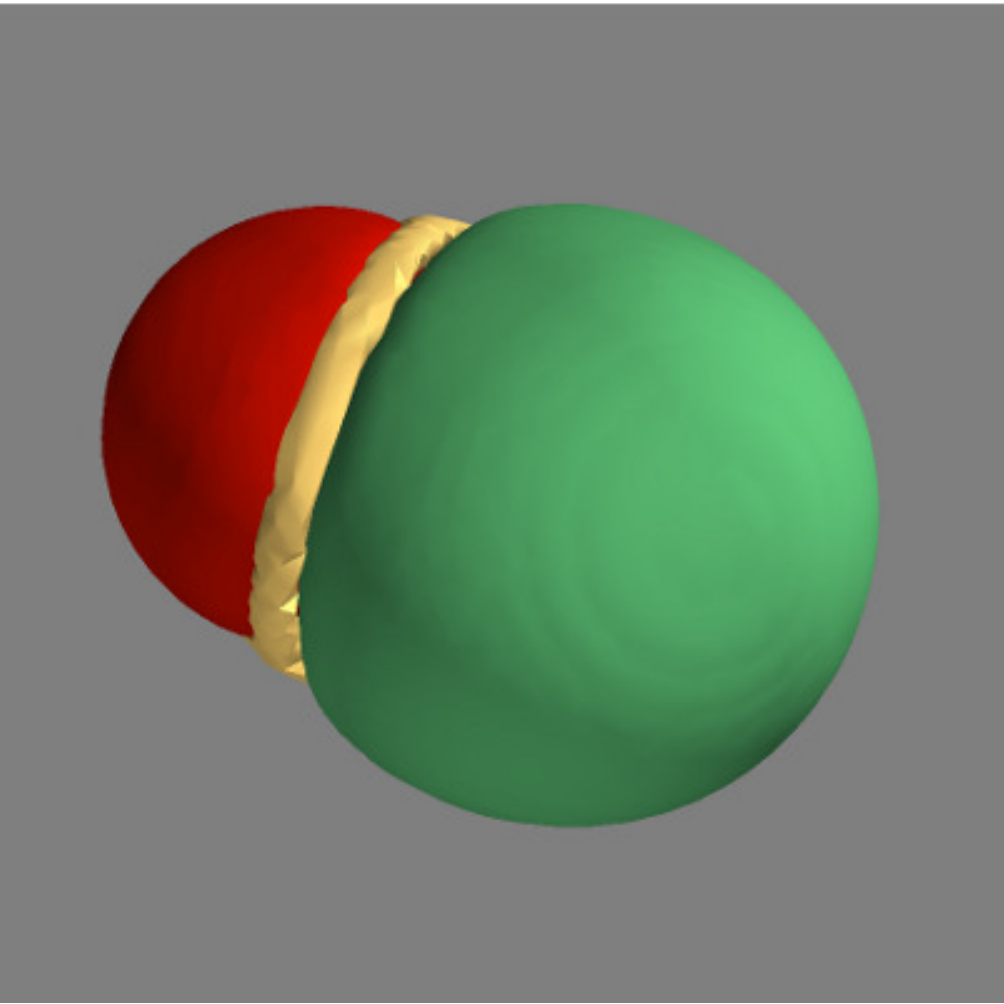}&
\includegraphics[scale=0.25]{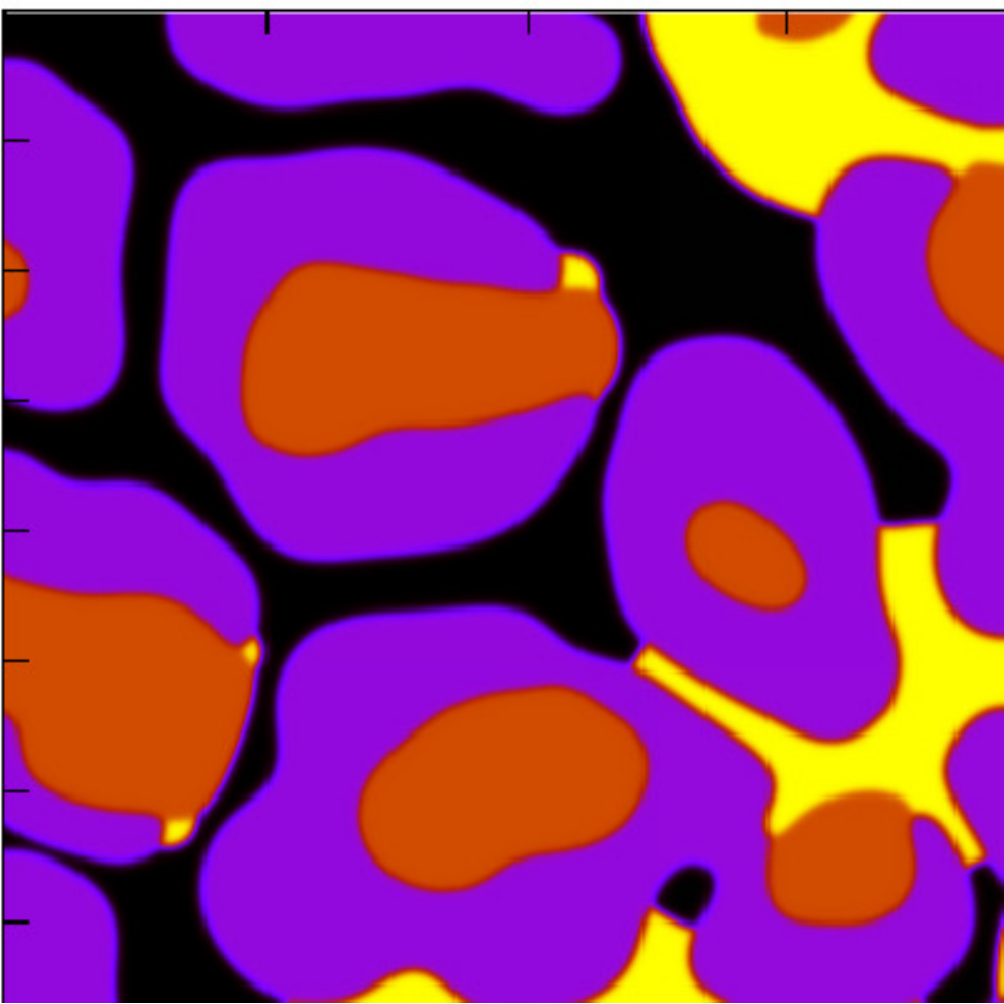}\\
(a)&(b)&(c)
\end{tabular}
\caption{
Examples of the nucleation events of the $\gamma$ phase in alloy 1 (a) and (b) and in alloy 2 (c). Red and yellow areas indicate $\beta$ and $\gamma$ phases, black (a) and gray (b) areas indicate the liquid phase, purple (a,c) and green (b) areas indicate the $\alpha$ phase.}
\end{centering}
\end{figure}

\begin{figure}
\label{Fig4}
\begin{centering}
\includegraphics[scale=0.3]{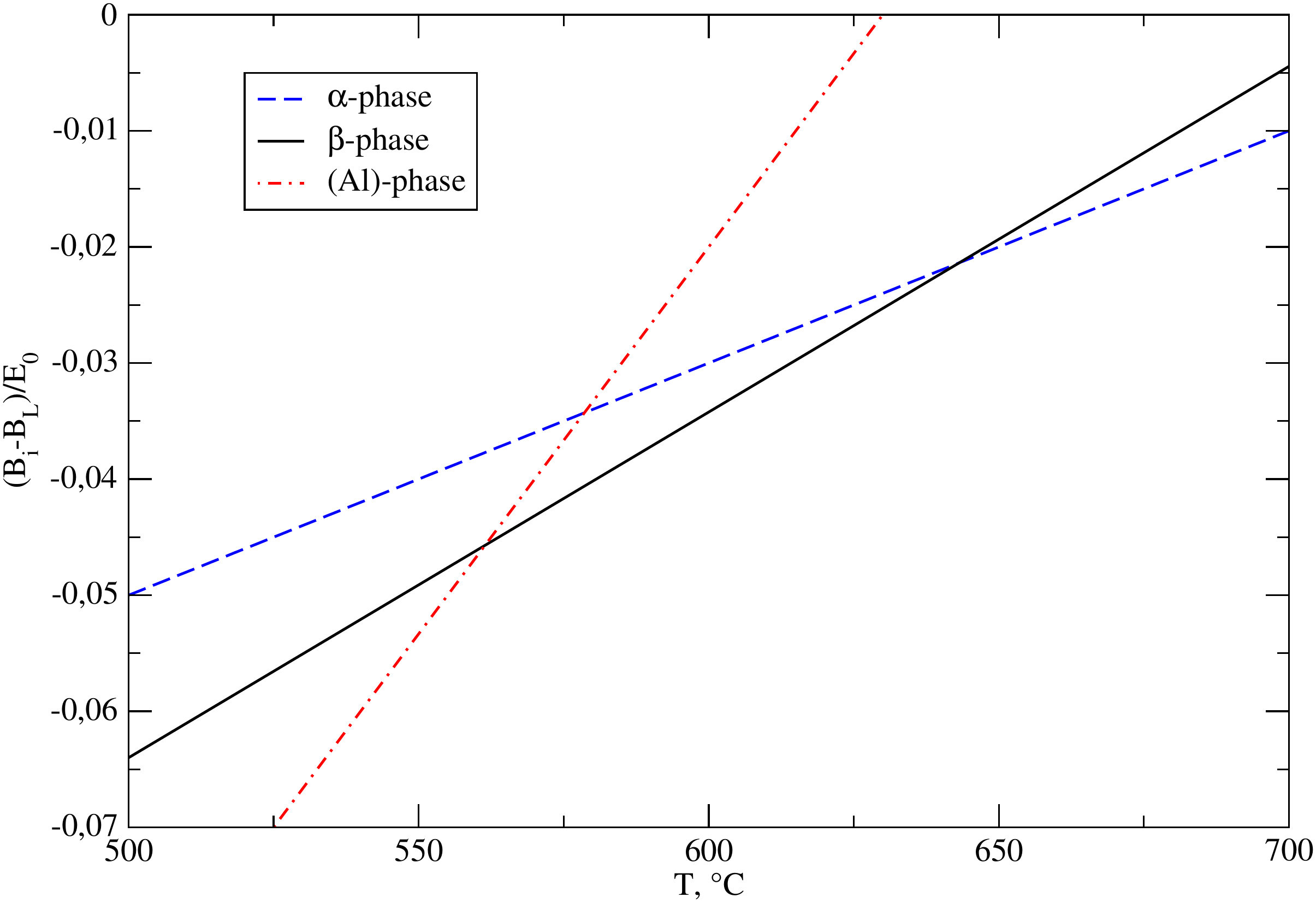}
\caption{
Scaled minimal energies of phases.}
\end{centering}
\end{figure}

The resulting microstructure and the concentration field of Ni at 520$^\circ$C are shown in Figure 5. This structure can be compared to the real structure of alloy 2 presented in Figure 6. In the phase-field model the functions serve for the stability of the solution and the absence of the third phase on individual interfaces. The explanation of the nucleation of a forth phase in triple points was done in the work \cite{Pogorelov13}. The authors show that in this points the presenence of small fluctuations of the forth phase is maximal and at the appropriate energetic conditions the nucleation and growth of new phase can occur. In our case the $\gamma$ phase has an enough small Gibbs free energy. Furthermore, the concentration of Al near the triple points is maximal that make the nucleation of the $\gamma$ phase very favorable. The spontaneous nucleation of the $\gamma$ phase in the triple points can be also controlled by the thermal noise. The number of the nucleation events in the system increases with 
increasing $\xi_0$. At  $\xi_0=0$, 0.15 and 0.3 the number of nucleation events was 2, 6 and 10, respectively.   Another interesting observation is that no nucleation of the third phase on dual solid/liquid interfaces can be observed even for larger $\xi_0$ unlike the simulations in alloy 1. This can be explained by the small energy difference between the phases.

\begin{figure}
\label{Fig5}
\begin{centering}
\begin{tabular}{ll}
\includegraphics[scale=0.2]{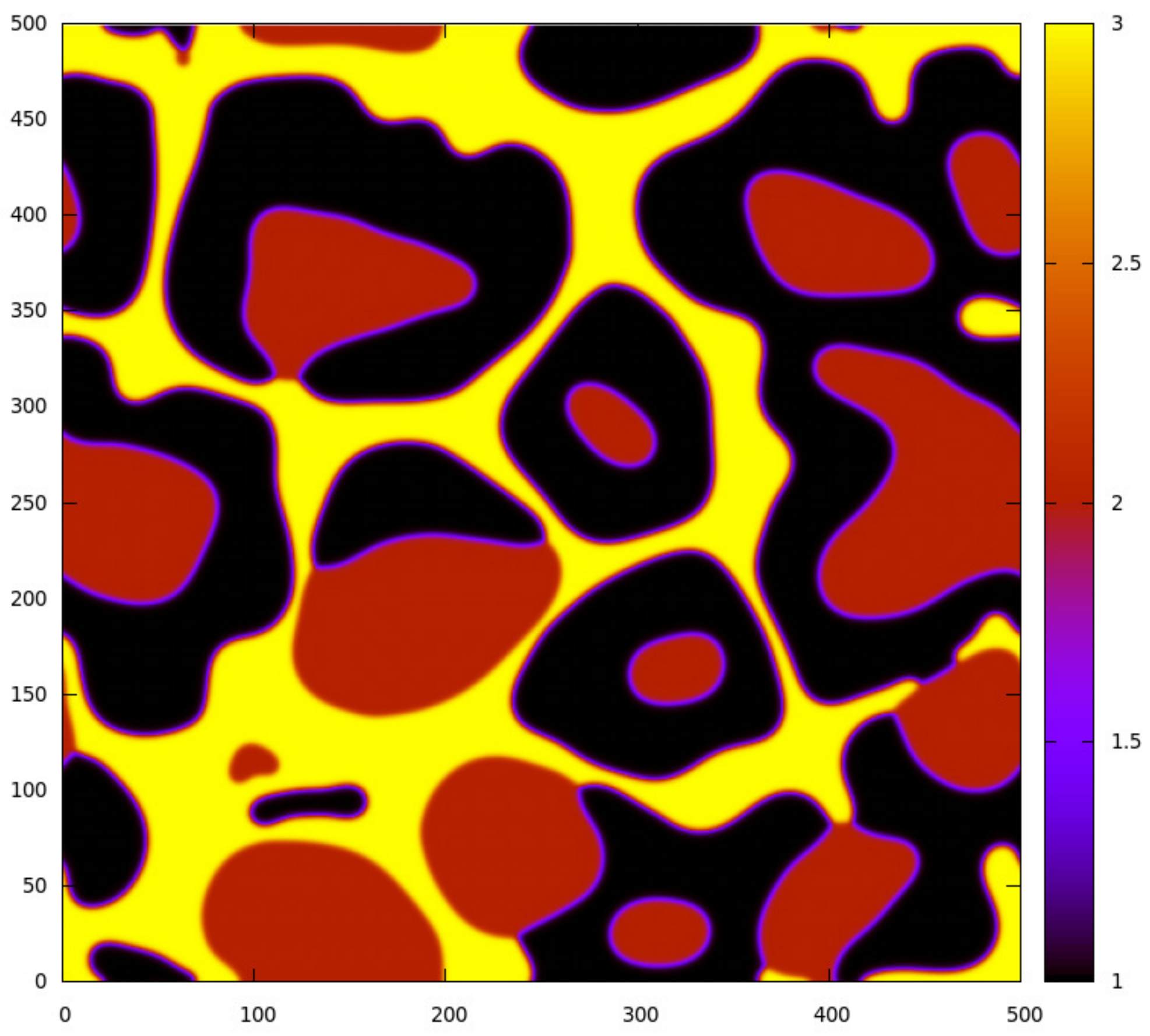}&
\includegraphics[scale=0.2]{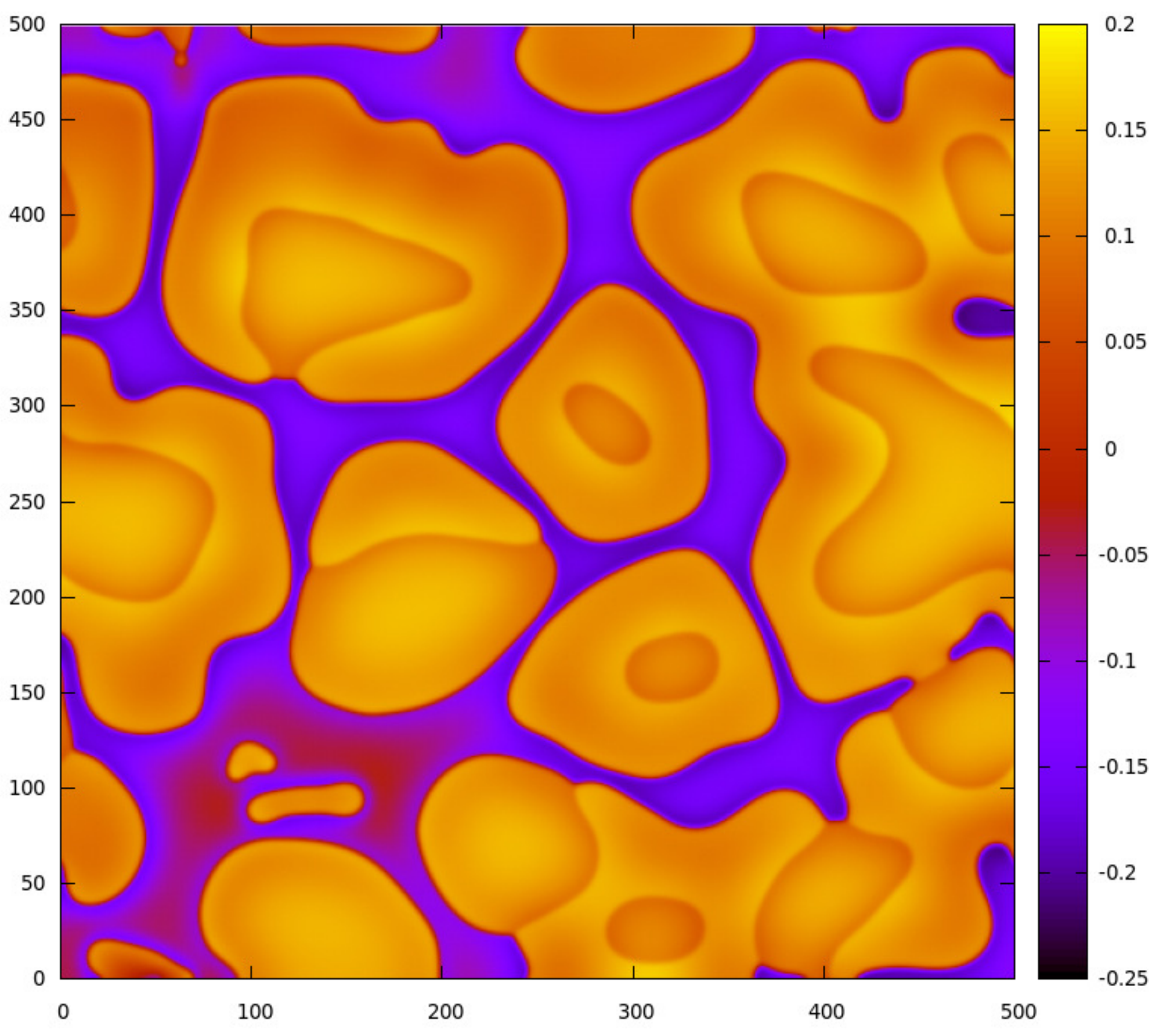}\\
(a)&(b)
\end{tabular}
\caption{
Simulated microstructure in alloy 2 (a) and the corresponding concentration field of Ni (b). Black, red and yellow areas indicate  $\alpha$, $\beta$ and $\gamma$ phases, respectively. }
\end{centering}
\end{figure}

The evolution of the phase fractions in alloy 2 is shown in Figure 7.
The comparison of the time evolution of the phase fractions for two simulated cases shows that the thermal noise increases the growth velocity of the $\gamma$ phase and reduces the time of the four-phase reaction. In a such a way it can influence the experimental DSC curves \cite{Kundin12b}. 

\begin{figure}
\label{Fig6}
\begin{centering}
\includegraphics[scale=0.25]{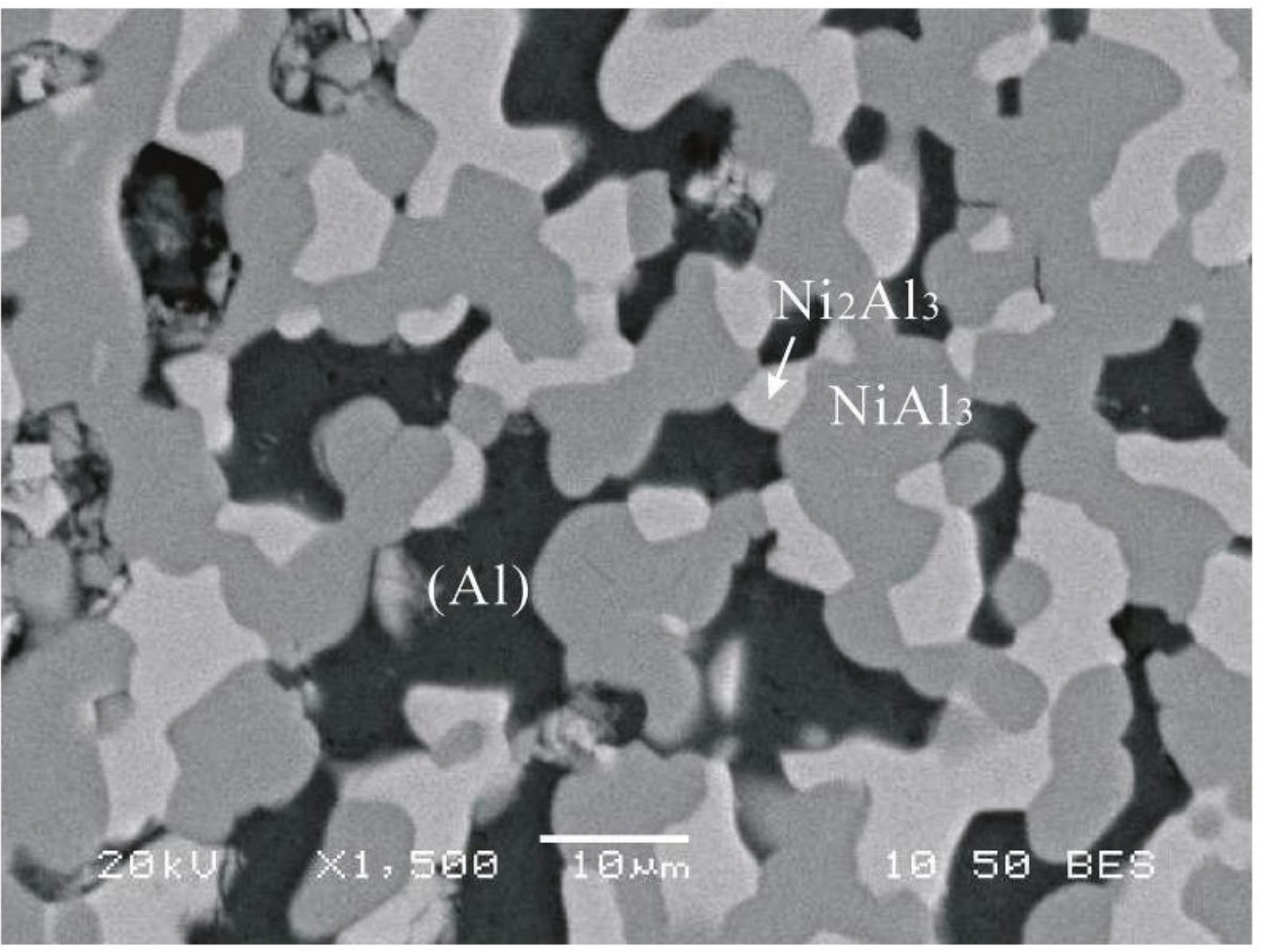}
\caption{
The backscatter micrograph of alloy Al75Cu6Ni19 equilibrated at 520$^\circ$C \cite{Schmid-Fetzer}. }
\end{centering}
\end{figure} 

\begin{figure}
\label{Fig7}
\begin{centering}
\includegraphics[scale=0.3]{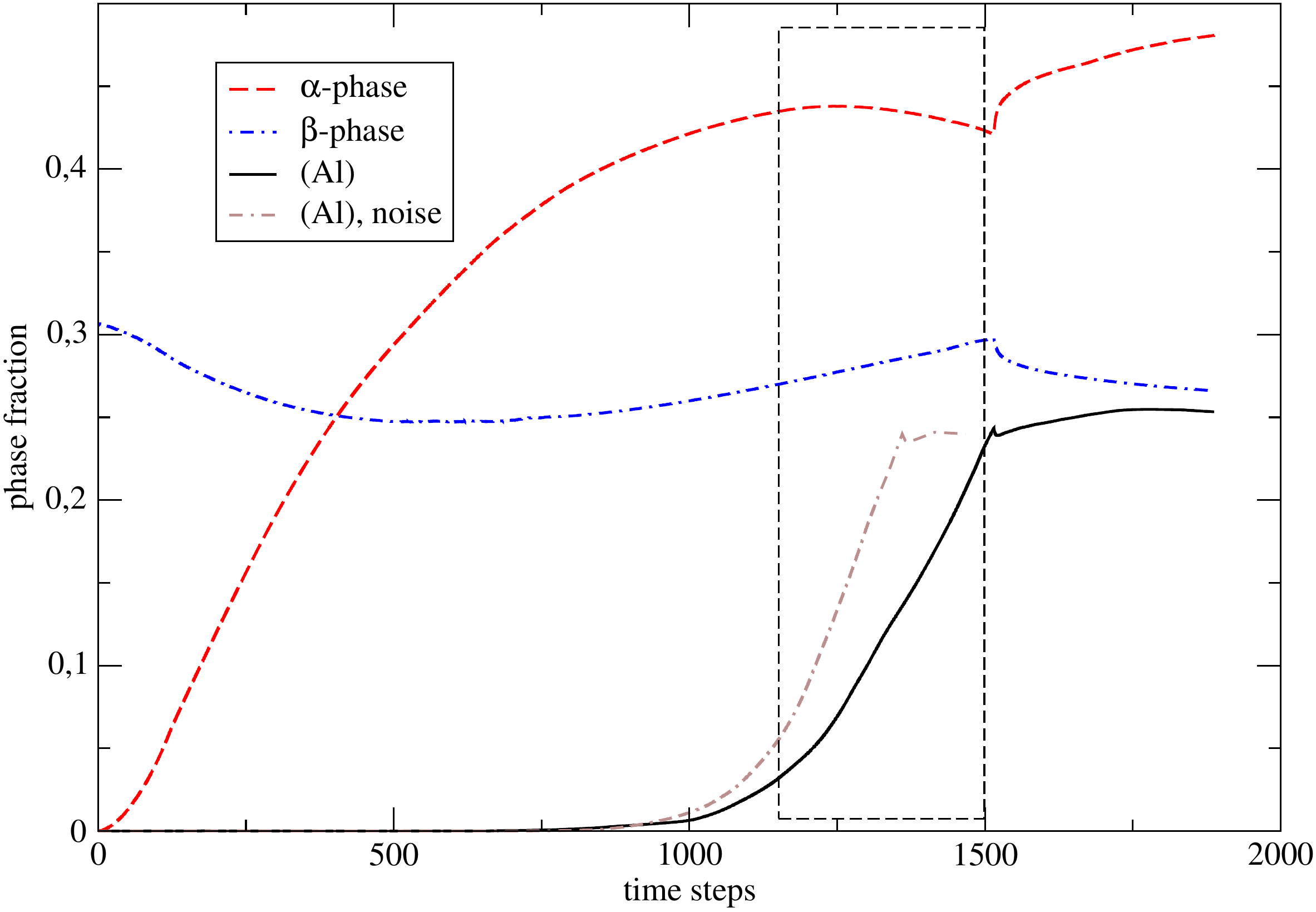}
\caption{The time evolution of phase fractions in alloy 2. A dashed box indicates the region of the four-phase reaction.}
\end{centering}
\end{figure}

In summary, on the basis of the general phase-field model for a $n$-dimensional phase-field space we have demonstrated the emergence of microstructure simulation in ternary alloys with four-phase interactions. It is theoretically clear, that the equilibrium parameters of phases used in the simulation and hence the initial alloy composition and also the process parameters such as the cooling velocity strongly influence the microstructure formation. Most importantly, the heterogeneous nucleation of new product phases can be done in the model by three ways. First way is the random insertion of a nucleus on the solid/liquid interface. The second one is the spontaneous nucleation of the forth phase in the triple points that is the organic property of the model and in consistence with the physical theory of the heterogeneous nucleation. The third way is the nucleation of a new phase on a dual interface by means of the special thermal noise term. Note that these terms are constructed using the derivatives of the 
model functions $g_i$ and can be seen as nature fluctuations of the thermodynamic driving forces. Of particular importance is that the 3D simulations give us the same nucleation effects, which we also observe in 2D simulations with only one difference that the forth phase nucleates on the triple lines. It can be predicted that in the case of a five-phase system we will obtain the spontaneous nucleation of the fifth phase in the quadropoints in the simulation.

\end{document}